\documentclass[twocolumn,twoside,slac_two]{revtex4}
\usepackage{graphicx}
\usepackage{fancyhdr}
\begin{document}

\title{Study of $D^{+}$ and $D^{+}_{s}$ Decay Properties at Belle}


\author{S. Lee}
\author{B. R. Ko}
\author{E. Won}
\affiliation{Department of Physics, Korea University, Seoul, Korea}
\author{and on behalf of the Belle Collaboration}

\begin{abstract}
We report the first observation of the doubly Cabibbo-suppressed decay, $D^{+}_{s} \to K^{+}K^{+}\pi^{-}$ and the significantly improved measurement of the doubly Cabibbo-suppressed decay, $D^{+} \to K^{+}\pi^{+}\pi^{-}$ using $\textrm{605 fb}^{-1}$ of data collected with the Belle detector at the KEKB asymmetric-energy $e^{+}e^{-}$ collider. The branching ratio with respect to its Cabibbo-favored counterpart are $\mathcal{B}(D^{+}_{s} \to K^{+}K^{+}\pi^{-})/\mathcal{B}(D^{+}_{s} \to K^{+}K^{-}\pi^{+}) = 0.229\pm0.028\pm0.012)\%$ and $\mathcal{B}(D^{+} \to K^{+}\pi^{+}\pi^{-})/\mathcal{B}(D^{+} \to K^{-}\pi^{+}\pi^{+}) = (0.569\pm0.018\pm0.014)\%$, where the first uncertainties are statistical and the second is systematic. 
We also report the improved measurement of $D^{+} \to K^{0}_{S}K^{+}$ and $D^{+}_{s} \to K^{0}_{S}\pi^{+}$ branching ratios using the same amount of data samples. The measured branching ratios with respect to the Cabibbo-favored modes are $\mathcal{B}(D^{+} \to K^{0}_{S}K^{+})/\mathcal{B}(D^{+} \to K^{0}_{S}\pi^{+}) = (0.1899\pm0.0011\pm0.0018)$ and $\mathcal{B}(D^{+}_{s} \to K^{0}_{S}\pi^{+})/\mathcal{B}(D^{+}_{s} \to K^{0}_{S}K^{+}) = (0.0803\pm0.0024\pm0.0012)$.

\end{abstract}

\maketitle

\thispagestyle{fancy}


\section{Introduction}

Cabibbo-suppressed (CS) and doubly Cabibbo-suppressed (DCS) decays play an important role in studies of charmed hadron dynamics. CS decays of nearly all the charmed hadrons have been observed, while DCS decays have been observed for only $D^{+}$ and $D^{0}$ mesons. The na\"ive expectation for the DCS decay rate is of the order of $\tan^{4}\theta_{C}$, where $\theta_{C}$ is the Cabibbo mixing angle \cite{PRL10_531}, or about $0.29\%$ relative to its Cabibbo-favored (CF) counterpart.\footnote{We use $\sin\theta_{C}=0.2257\pm0.0010$ as given in \cite{PLB667_1} for the numerical estimate.}

One expects that the branching ratio of $D^{+} \to K^{+}\pi^{+}\pi^{-}$ is about $2\tan^{4}\theta_{C}$ relative to its CF counterpart since the phase space for $D^{+} \to K^{-}\pi^{+}\pi^{+}$ is suppressed due to the two identical pions in the final state.\footnote{Throughout this paper the charge-conjugate state is implied unless stated otherwise.} This expectation is consistent with current experimental results \cite{PLB667_1}. Therefore, we also expect the branching ratio of $D^{+}_{s} \to K^{+}K^{+}\pi^{-}$ is about $1/2\tan^{4}\theta_{C}$ relative to its CF counterpart. Lipkin \cite{NPBPS115_117} argues that SU(3) flavor symmetry\footnote{SU(3) flavor symmetry implies invariance under the transformations $K^{\pm} \leftrightarrow \pi^{\pm}$, $D^{\pm} \leftrightarrow D^{\pm}_{s}$.} implies
\begin{eqnarray*}
\frac{\mathcal{B}(D^{+}_{s} \to K^{+}K^{+}\pi^{-})}{\mathcal{B}(D^{+}_{s} \to K^{+}K^{-}\pi^{+})}
\frac{\mathcal{B}(D^{+} \to K^{+}\pi^{+}\pi^{-})}{\mathcal{B}(D^{+} \to K^{-}\pi^{+}\pi^{+})}
& = & \tan^{8}\theta_{C}
\end{eqnarray*}
where differences in the phase space for CF and DCS decay modes cancel in the ratios. The above relation does not take into account possible SU(3) breaking effects that could arise due to resonant intermediate states in the three-body final states considered here \cite{NPBPS115_117}.

In addition, decays of charmed mesons play important role in understanding the sources of the SU(3) flavor symmetry breaking effects \cite{PRD77_114020}. Such breaking effects can originate from strong final-state interactions or interference between same final states. In particular, $D^{+} \to \overline{K}^{0}K^{+}$ and $D^{+}_{s} \to K^{0}\pi^{+}$ are CS decays with the color-favored tree, annihilation and penguin diagrams. For $D^{+}$ decays, the branching ratio $\mathcal{B}(D^{+} \to \overline{K}^{0}K^{+})/\mathcal{B}(D^{+} \to \overline{K}^{0}\pi^{+})$ deviates from the na\"ive expectation of $\tan^{2}\theta_{C}$ \cite{PLB667_1}, due to a destructive interference between color-favored and color-suppressed amplitude in $D^{+} \to \overline{K}^{0}\pi^{+}$ \cite{PLB89_111}. However, converting experimental measurements including $K^{0}_{S}$ branching ratios to those of involving $\overline{K}^{0}$ is not straightforward due to the fact that one must take into account the interference between DCS decay and CF decay modes where the interference phase is unknown \cite{PLB349_363}. In $D^{+}_{s}$ decays to $\overline{K}^{0}K^{+}$ and $K^{0}\pi^{+}$ final states, the ratio of CS decay to that of the corresponding CF decay may be larger than $\tan^{2}\theta_{C}$, since $D^{+}_{s}\to\overline{K}^{0}K^{+}$ decay mode is a CF but color-suppressed decay mode. Precise branching ratio measurements of CS and CF charm meson decay modes can thus help in understanding the underlying dynamics of these decays.

The data used in the analysis were recorded at the $\Upsilon(4S)$ resonance with the Belle detector \cite{NIMA479_117}  at the $e^{+}e^{-}$ asymmetric-energy collider KEKB \cite{NIMA499_1}. It corresponds to an integrated luminosity of $\textrm{605 fb}^{-1}$.

\section{First observation of DCS decay in $D_{s}^{+}\to K^{+}K^{+}\pi^{-}$ and improved measurement of $D^{+}\to K^{+}\pi^{+}\pi^{-}$}

We require that the charged tracks originate from the vicinity of the interaction point with impact parameters in the beam direction ($z$-axis) and perpendicular to it of less than 4 cm and 2 cm, respectively. All charged tracks are required to have at least two associated hits in the silicon vertex detector \cite{NIMA560_1}, both in the $z$ and radial directions, to assure good spatial resolution on the $D$ mesons' decay vertices. The decay vertex is formed by fitting the three charged tracks to a common vertex and requiring a confidence level (C.L.) greater than 0.1\%. Charged kaons and pions are identified requiring the ratio of particle identification likelihoods, $\mathcal{L}_{K}/(\mathcal{L}_{K}+\mathcal{L}_{\pi})$, constructed using information from the central drift chamber, time-of-flight counters, and aerogel Cherenkov counters \cite{NIMA494_402}, to be larger or smaller than 0.6, respectively. In addition, we require that the scaled momentum of the charmed meson candidate $x_{p}=p^{*}/\sqrt{0.25E^{2}_{\mathrm{CM}}-M^{2}}$ be greater than 0.5 to suppress combinatorial background as well as $D$ mesons produced in $B$ meson decays. Here $p^{*}$ is the charmed meson momentum and $E_{\mathrm{CM}}$ is the total $e^{+}e^{-}$ collision energy calculated in the center-of-mass frame. 

In order to suppress the high background level, we apply further selection criteria, which are optimized using real data samples since there are some discrepancies between the Monte Carlo (MC) simulation \cite{MC} and the data in the relevant distributions. We use 10\% of the data samples for optimization and the remaining 90\% for the measurement to avoid a possible bias when the same samples are used for both optimization and the measurement. Assuming no signal in the DCS decay channel, we maximize $\mathcal{N}_{S}/\sqrt{\mathcal{N}_{B}}$, where $\mathcal{N}_{S}$ is the CF signal yield which has similar properties to the DCS signal and $\mathcal{N}_{B}$ is the background yield from the sideband regions in the DCS sample.

\begin{figure}[b]
\centering
\includegraphics[width=80mm]{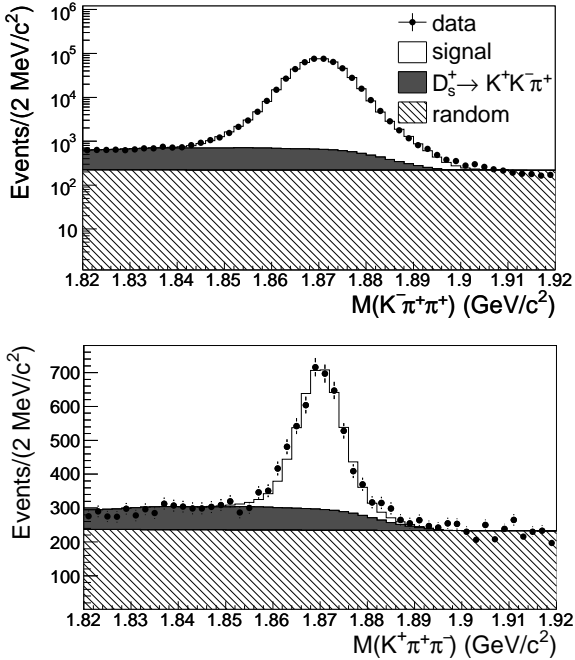}
\caption{Distributions of $M(K^{-}\pi^{+}\pi^{+})$ (top) and $M(K^{+}\pi^{+}\pi^{-})$ (bottom). The $K^{-}\pi^{+}\pi^{+}$ distribution is shown on a semilogarithmic scale to make the small background visible. Points with error bars show the data and histograms show the results of the fits described in the text. Signal, $D^{+}_{s}\to K^{+}K^{-}\pi^{+}$ background, and random combinatorial background components are also shown.} \label{fig_02.eps}
\end{figure}

\begin{figure}[t]
\centering
\includegraphics[width=80mm]{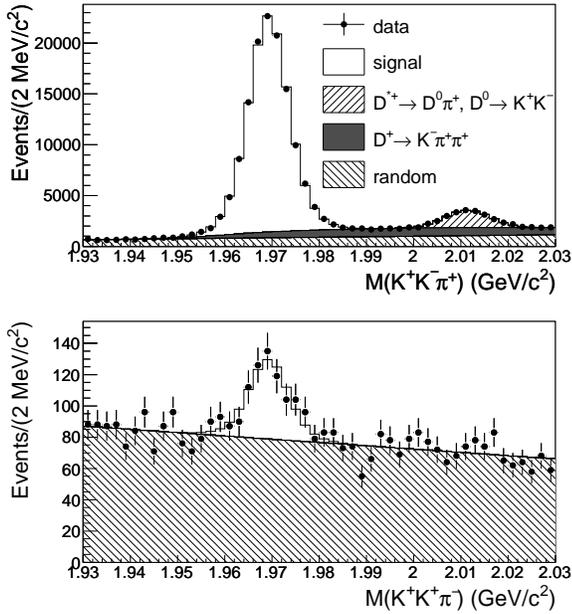}
\caption{Distribution of $M(K^{+}K^{-}\pi^{+})$ (top) and $M(K^{+}K^{+}\pi^{-})$ (bottom). Points with error bars show the data and histograms show the results of the fits described in the text. Signal, $D^{*+}$ background ($D^{*+}\to D^{0}\pi^{+}$ where $D^{0}\to K^{+}K^{-}$), $D^{+}\to K^{-}\pi^{+}\pi^{+}$ background, and random combinatorial background components are also shown.} \label{fig_03.eps}
\end{figure}

One of the selections related to the finite lifetime of charmed hadrons is the reduced $\chi^{2}$ ($\chi^{2}$/d.o.f) for the hypothesis that the candidate tracks for the charmed meson decay products arise from the primary vertex. The primary vertex is obtained as the most probable point of intersection of the meson's momentum vector and the $e^{+}e^{-}$ interaction region. Because of the finite lifetime of $D^{+}$ and $D^{+}_{s}$ mesons their daughter tracks are not likely to be compatible with the primary vertex. The second requirement uses the angle between the charmed meson momentum vector, as reconstructed from the daughter tracks, and the vector joining its production and decay vertices. In an ideal case without resolution the two vectors would be parallel for the signal. The reduced $\chi^{2}$ is required to be greater than 25 for $D^{+}$ and 5 for $D^{+}_{s}$ candidates and the angle is required to be less than $1^{\circ}$ for $D^{+}$ and $2^{\circ}$ for $D^{+}_{s}$ candidates. Tighter requirements on charged kaon identification ($>0.9$) and $x_{p}$ ($>0.7$) are also chosen for the final selection, which improves the signal sensitivity. After the additional and tighter selection requirements described above, 9.57\% of $D^{+}$ and 10.71\% of $D^{+}_{s}$ CF signal and 0.06\% of $D^{+}$ and 0.24\% of $D^{+}_{s}$ DCS background events are retained. In order to minimize systematic effects we choose the same selection criteria for both DCS and CF decay channels. The $K\pi\pi$ and $KK\pi$ invariant mass distributions after the final selections are shown in Figures \ref{fig_02.eps} and \ref{fig_03.eps} together with signal and background parametrizations. A clear signal is observed in both DCS decay mass distributions.

The $K\pi\pi$ and $KK\pi$ invariant mass distributions are fitted using the binned maximum likelihood method. In all cases the signal probability density function (PDF) is parametrized using two Gaussians with the same central value. Because of $K/\pi$ misidentification the following reflection backgrounds appear in the mass distributions. In $D^{+}\to K^{-}\pi^{+}\pi^{+}$ (CF) and $K^{+}\pi^{+}\pi^{-}$ (DCS) decays there is a contribution from misidentified $D^{+}_{s}\to K^{+}K^{-}\pi^{+}$ decays; in $D^{+}_{s}\to K^{+}K^{-}\pi^{+}$ (DCS) decay there is a contribution from misidentified $D^{+}\to K^{+}\pi^{+}\pi^{-}$ decay. The PDFs for the refection backgrounds are determined from real data by assigning the nominal pion (kaon) mass to a kaon (pion) track. The magnitude of each of the reflection background contributions is a free parameter in the fit. For the DCS $D^{+}_{s}$ channel, the $D^{+}\to K^{+}\pi^{+}\pi^{-}$ contribution is not incorporated in the fit since it is not significant, but its effect is included as a systematic uncertainty due to fitting listed in Table \ref{DCS_Systematics}. The $D^{*+}$ contributions ($D^{*+}\to D^{0}\pi^{+}$ with $D^{0}\to K^{+}K^{-}$) in the CF $D^{+}_{s}$ channel is also incorporated in the CF $D^{+}_{s}$ fit as an independent Gaussian component. A linear function is used for the random combinatorial background for all channels. All signal and background parameters for the CF channels are floated. For the DCS channels the mass, width, and ratio of the two signal Gaussians are fixed to the values obtained from the fits to distributions of CF decays. Signal and background yields are left free in the fit. From the results of the fits, shown in Figures \ref{fig_02.eps} and \ref{fig_03.eps}, we extract the signal yield for each channel, listed together with the corresponding branching ratios in Table \ref{BR}.

\begin{table}[b]
\begin{center}
\caption{Measured branching ratios. $\mathcal{B}_{rel}$ is the branching ratio relative to $D^{+}\to K^{-}\pi^{+}\pi^{+}$ for the $D^{+}$ modes and $D^{+}_{s}\to K^{+}K^{-}\pi^{+}$ for the $D^{+}_{s}$ modes. The uncertainties in the branching ratios are statistical and systematic.}
\begin{tabular}{lcc}
\hline\hline
Decay Mode & $\mathcal{N}_{signal}$ & $\mathcal{B}_{rel}$ (\%) \\
\hline
$D^{+}\to K^{+}\pi^{+}\pi^{-}$ & $2637.7\pm 84.4$ & $0.569\pm 0.018 \pm 0.014$ \\
$D^{+}\to K^{-}\pi^{+}\pi^{+}$ & $482702 \pm 727$ & $100$ \\
$D^{+}_{s}\to K^{+}K^{+}\pi^{-}$ & $281.4 \pm 33.8$ & $0.229\pm 0.028\pm 0.012$ \\
$D^{+}_{s}\to K^{+}K^{-}\pi^{+}$ & $118127\pm 452$ & $100$ \\
\hline\hline
\end{tabular}
\label{BR}
\end{center}
\end{table}

The statistical significance of the $D^{+}_{s}\to K^{+}K^{+}\pi^{-}$ signal is calculated using $-2\ln(\mathcal{L}_{b}/\mathcal{L}_{s+b})$ where $\mathcal{L}_{b}$ and $\mathcal{L}_{s+b}$ are the likelihood values of the fit, without and with the signal PDF included, respectively.  We find $-2\ln(\mathcal{L}_{b}/\mathcal{L}_{s+b}) = 83.2$ with 1 degree of freedom used to describe the DCS signal yield; we obtain a statistical significance corresponding to 9.1 standard deviations.

In addition to the backgrounds mentioned above there is also the possibility of double misidentification leading to contributions from CF events to the DCS sample. MC simulation shows that such a contribution is flat in the invariant mass distribution and is hence included in the combinatorial background description.

\begin{table}[b]
\begin{center}
\caption{Relative systematic uncertainties in percent, where $\sigma_{\mathcal{B}_{rel}(D^{+})}$ and $\sigma_{\mathcal{B}_{rel}(D^{+}_{s})}$ are systematic uncertainties for the branching ratio of $D^{+}$ and $D^{+}_{s}$ DCS decays relative to their CF counterparts.}
\begin{tabular}{lcc}
\hline\hline
Source & $\sigma_{\mathcal{B}_{rel}(D^{+})}$ & $\sigma_{\mathcal{B}_{rel}(D^{+}_{s})}$ \\
\hline
Fitting & 1.9 & 4.2 \\
MC  statistics & 0.8 & 1.0 \\
Reconstruction efficiency & 1.5 & 3.1 \\
\hline
Total & 2.5 & 5.3 \\
\hline\hline
\end{tabular}
\label{DCS_Systematics}
\end{center}
\end{table}

\begin{table*}[t]
\begin{center}
\caption{Absolute branching fraction for each decay mode and comparisons with previous measurements. The first uncertainties shown in the second column are the total uncertainties of our results, and the second are the uncertainties in the average CF branching fractions used for normalization \cite{PLB667_1}.}
\begin{tabular}{lcc}
\hline\hline
Branching fraction & Belle & World average \cite{PLB667_1} \\
\hline
$\mathcal{B}(D^{+}\to K^{+}\pi^{+}\pi^{-})$ & $(5.2\pm 0.2\pm 0.1)\times 10^{-4}$ & $(6.2\pm 0.7)\times 10^{-4}$ \\
$\mathcal{B}(D^{+}_{s}\to K^{+}K^{+}\pi^{-})$ & $(1.3\pm 0.2\pm 0.1)\times 10^{-4}$ & $(2.9\pm 1.1)\times 10^{-4}$ \\
\hline\hline
\end{tabular}
\label{DCS_Result}
\end{center}
\end{table*}

The final states in this study have resonant substructure that can affect the reconstruction efficiency. The resonances are relatively well known for the decay modes other than $D^{+}_{s}\to K^{+}K^{+}\pi^{-}$. We used a coherent mixture of resonant contributions according to \cite{PRD48_56} to generate $D^{+}\to K^{-}\pi^{+}\pi^{+}$ decays and calculate the reconstruction efficiency. For the $D^{+}\to K^{+}\pi^{+}\pi^{-}$ and $D^{+}_{s}\to K^{+}K^{-}\pi^{+}$ decays we used an incoherent mixture of intermediate states \cite{PLB667_1}. Subsequently we varied the contributions of individual intermediate states in a correlated manner, within the uncertainties of the measured branching fractions. The efficiency calculated from the modified MC sample differs from the original one by 1.5\% and 2.0\% for the $D^{+}\to K^{+}\pi^{+}\pi^{-}$ and $D^{+}_{s}\to K^{+}K^{-}\pi^{+}$ decays, respectively, and the difference was included in the systematic uncertainty of the result. $D^{+}_{s}\to K^{+}K^{+}\pi^{-}$ decays were generated according to phase space. For comparison, signal events were generated assuming either $K^{*0}(802)K^{+}$ or $K^{*0}(1430)K^{+}$ intermediate states. The largest relative difference in the efficiency (2.4\%) was included as a part of the systematic uncertainty. Ratios of reconstruction efficiencies for DCS and CF decays are found to be $1.042\pm 0.008\pm 0.016$ and $0.963\pm 0.010\pm 0.030$ for $D^{+}$ and $D^{+}_{s}$ decays, respectively, where the first uncertainty is due to the finite MC simulation statistics and the second is the uncertainty in the resonant structure of the final states.

With the efficiencies estimated above, we measure the inclusive branching ratios of DCS decays relative to their CF counterparts summarized in Table \ref{BR}. The product of the branching ratios for the two DCS decay modes is found to be $\frac{\mathcal{B}(D^{+}_{s}\to K^{+}K^{+}\pi^{-})}{\mathcal{B}(D^{+}_{s}\to K^{+}K^{-}\pi^{+})}\frac{\mathcal{B}(D^{+}\to K^{+}\pi^{+}\pi^{-})}{\mathcal{B}(D^{+}\to K^{-}\pi^{+}\pi^{+})} = (1.57\pm 0.21) \tan^{8}\theta_{C}$, where the error is the total uncertainty.

Several sources of systematic uncertainty cancel in the branching ratio calculation due to the similar kinematics of CF and DCS decays (for example, uncertainties in the tracking efficiencies and particle identification, since the momenta of the final state tracks are almost identical). The stability of the branching ratios against the variation of the selection criteria was studied and we observed no changes greater than the expected statistical fluctuations. The systematic uncertainties due to the variation of the fit parameters are 1.9\% for $D^{+}$ and 4.2\% for $D^{+}_{s}$ branching ratios measurements, respectively. Table \ref{DCS_Systematics} summarizes the systematic uncertainties in the measurements of the branching ratios.

Using the world average values $\mathcal{B}(D^{+}\to K^{-}\pi^{+}\pi^{+}) = (9.22\pm 0.21)\%$ and $\mathcal{B}(D^{+}_{s}\to K^{+}K^{-}\pi^{+}) = (5.50\pm 0.28)\%$ \cite{PLB667_1}, we obtain the absolute branching fraction for each DCS decay channel. Table \ref{DCS_Result} shows the comparison between previous results and this work, published in \cite{PRL102_221802}.

\section{Precise measurement of $D^{+}\to K^{0}_{S}K^{+}$ and $D^{+}_{s}\to K^{0}_{S}\pi^{+}$ branching ratios}

We require similar selection criteria described in previous section. In addition, we require that $K^{0}_{S}$ candidates have the invariant mass within $\pm 9$ MeV/$c^{2}$ from its nominal $K^{0}_{S}$ mass \cite{PLB667_1}. Further requirements are also imposed to improve the quality of $K^{0}_{S}$ candidates \cite{PRD72_012004}. $D^{+}$ and $D^{+}_{s}$ candidates are reconstructed using a $K^{0}_{S}$ and a charged track. The decay vertex is formed by fitting the $K^{0}_{S}$ and the track to a common vertex and requiring a confidence level greater than 0.1\%. In order to remove peaking backgrounds from $D^{+}_{(s)}\to \pi^{+}\pi^{+}\pi^{-}$ and $K^{+}\pi^{+}\pi^{-}$ decay modes, we compute the reduced $\chi^{2}$ of the vertex assuming that two pions from $K^{0}_{S}$ and the charm daughter track arises from a single vertex. We require the reduced $\chi^{2}$ to be greater than 10. To completely remove $D^{+}$ and $D^{+}_{s}$ mesons produced in $B$ meson decays, we require $p^{*}$ to be greater than 2.6 GeV/$c$. Reconstruction efficiencies are found to be 16.6\% for $D^{+}$ and 18.0\% for $D^{+}_{s}$ in $K^{0}_{S}K^{+}$ final states, and 20.6\% for $D^{+}$ and 22.4\% for $D^{+}_{s}$ in $K^{0}_{S}\pi^{+}$ final states, at this stage.

\begin{figure}[t]
\centering
\mbox{\includegraphics[width=80mm]{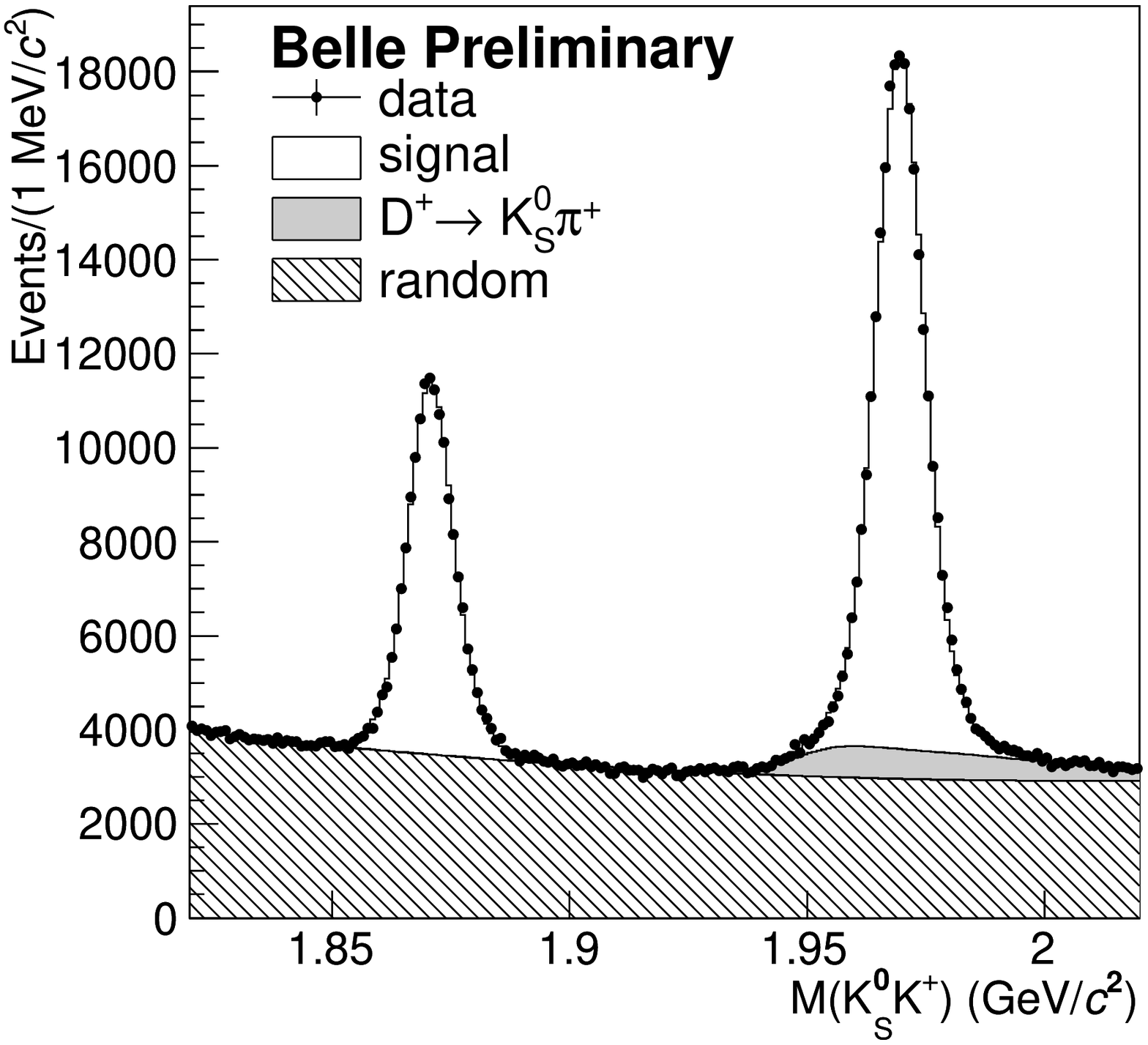}}
\mbox{\includegraphics[width=80mm]{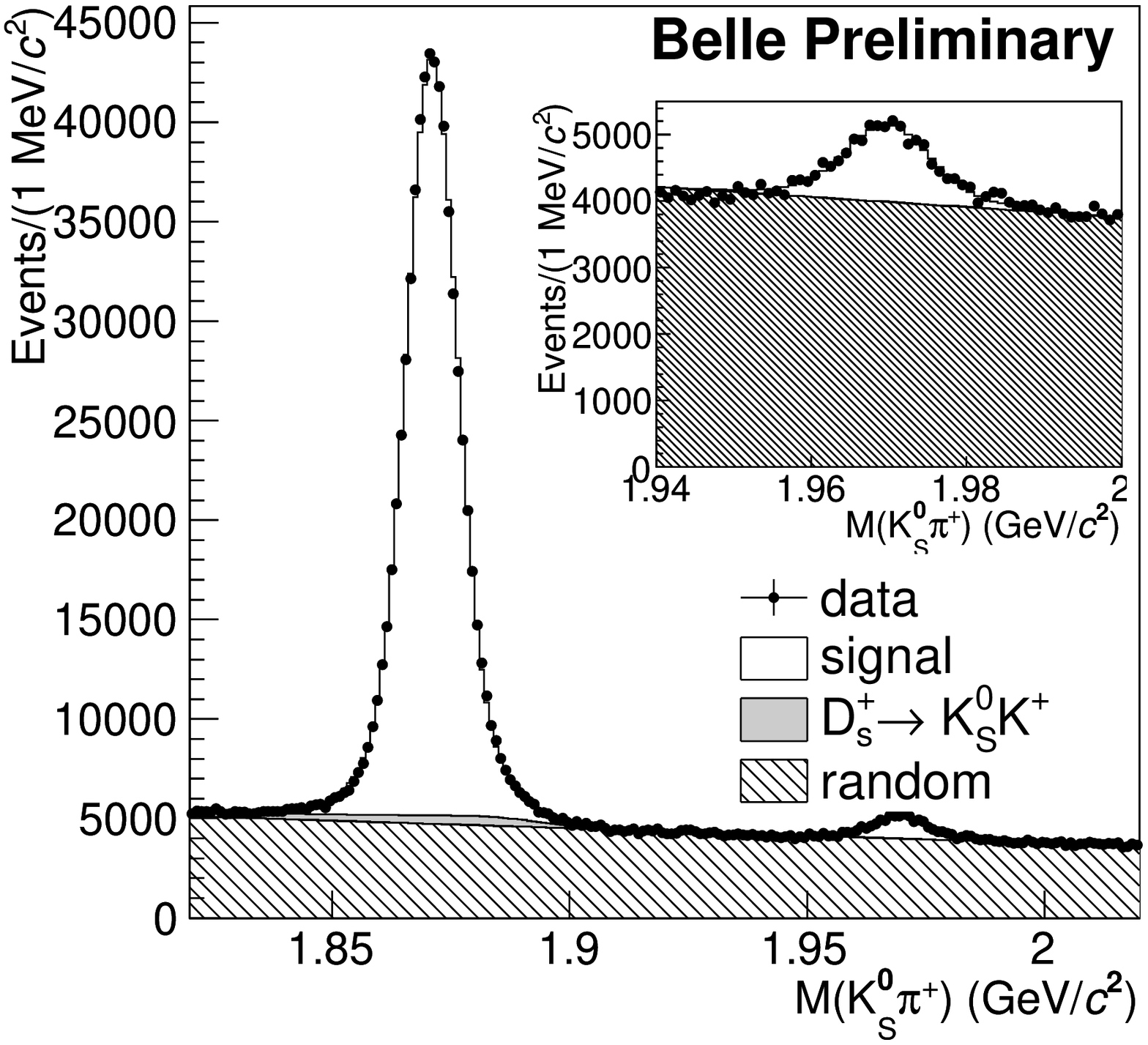}}
\caption{Invariant mass distribution of selected $K^{0}_{S}K^{+}$ (top) and $K^{0}_{S}\pi^{+}$ (bottom) pairs. Points with error bars show the data and histograms show the results of the fits described in the text. Signals, $D^{+}\to K^{0}_{S}\pi^{+}$ (top) and  $D^{+}_{s}\to K^{0}_{S}K^{+}$ (bottom) backgrounds, and random combinatorial background components are also shown. Inset of bottom is an enlarged view of $D^{+}_{s}$ region.} \label{fig_04.eps}
\end{figure}

Highly asymmetrical $K^{0}_{S}h^{+}$ pairs that have invariant mass  close to $D^{+}_{(s)}$ mass region are more likely to be background than signal. The selection optimization is made on asymmetry, $\mathcal{A}\equiv \left| p_{K^{0}_{S}} - p_{h^{+}}\right| / \left| p_{K^{0}_{S}} + p_{h^{+}}\right|$ to reject background candidates, where momentum is calculated in the laboratory frame and $h^{+}$ refers to either $K^{+}$ or $\pi^{+}$. We use the off-resonance data sample for optimization and the on-resonance data for the measurements as well as described in previous section. 

We perform a tuning of MC simulated events \cite{MC} intended mainly for the parametrization of the peaking background under the signal. This background is a consequence of particle misidentification and is to be discussed in more details later. The $\pi^{+}$ ($K^{+}$) momentum and its resolution are tuned with $D^{0}\to \pi^{+}\pi^{-}$ ($D^{0}\to K^{+}K^{-}$) data sample. For $K^{0}_{S}$ momentum scale and resolution tuning, $D^{+}\to K^{0}_{S}\pi^{+}$ data sample is used. The tuning method is applied to $K^{0}_{S}K^{+}$ final state simulated data and compared to the real data. The agreement between the simulated and the real data is significantly improved.

In the branching ratio measurements, there is a peaking background due to particle misidentification. In $D^{+}_{s}\to K^{0}_{S}K^{+}$ mass window, a peaking structure appears from $D^{+}\to K^{0}_{S}\pi^{+}$ decays when $\pi^{+}$ is misidentified as $K^{+}$. A similar peaking structure in $D^{+}\to K^{0}_{S}\pi^{+}$ mass window appears due to the misidentification in $D^{+}_{s}\to K^{0}_{S}K^{+}$ decays. The shapes and the yields of these peaking backgrounds are obtained from the simulated event samples. The simulated shape and normalization of the peaking background is checked by comparison to the invariant mass distributions of selected $K^{0}_{S}K^{+}$ ($K^{0}_{S}\pi^{+}$) pairs with $K^{+}$($\pi^{+}$) mass swapped by the $\pi^{+}$($K^{+}$) mass. The comparison shows that the simulated peaking background of tuned sample correctly describes these components and that the mentioned misidentification is indeed the only contribution above the structureless combinatorial background. Possible uncertainties in the misidentification probabilities are considered as a source of systematic uncertainty.

\begin{table}[t]
\begin{center}
\caption{The extracted signal yields (preliminary) from the fits to data and corresponding signal efficiencies ($\epsilon$) from the simulated events of signal modes. The uncertainties are statistical only.}
\begin{tabular}{lcc}
\hline\hline
Decay modes & Yields & $\epsilon$ (\%) \\
\hline
$D^{+}\to K^{0}_{S}K^{+}$ & $100855\pm 651$ & $12.69 \pm 0.01$ \\
$D^{+}_{s}\to K^{0}_{S}K^{+}$ & $204093\pm 768$ & $13.53\pm 0.01$ \\
$D^{+}\to K^{0}_{S}\pi^{+}$ & $566285\pm 1162$ & $14.19\pm 0.01$ \\
$D^{+}_{s}\to K^{0}_{S}\pi^{+}$ & $17583\pm 481$ & $15.35\pm 0.01$ \\
\hline\hline
\end{tabular}
\label{CS_Yields}
\end{center}
\end{table}

\begin{table}[b]
\begin{center}
\caption{Relative systematic uncertainties in percent (preliminary), where $\sigma_{R(D^{+})}$ and $\sigma_{R(D^{+}_{s})}$ are systematic uncertainties for branching ratios of $D^{+}$ and $D^{+}_{s}$ decays. Sources include particle identification (PID), fit methods, peaking background, and $D^{+}_{s}$ signal PDF.}
\begin{tabular}{lcc}
\hline\hline
Source & $\sigma_{R(D^{+})}$ (\%) & $\sigma_{R(D^{+}_{s})}$ (\%) \\
\hline
PID & 0.90 & 0.90 \\
Fit methods & 0.24 & 0.87 \\
peaking background & 0.16 & 0.62 \\
$D^{+}_{s}$ signal PDF & - & 0.37 \\
\hline
Total & 0.95 & 1.45 \\
\hline\hline
\end{tabular}
\label{CS_Systematics}
\end{center}
\end{table}

The $K^{0}_{S}K^{+}$ and $K^{0}_{S}\pi^{+}$ invariant mass distributions after the final selections are shown in Figure \ref{fig_04.eps} with signal and background parametrizations. Enhanced signals of CF and CS decays are observed in both distributions. The $K^{0}_{S}K^{+}$ and $K^{0}_{S}\pi^{+}$ invariant mass distributions are fitted using the binned maximum likelihood method. In all cases the signal PDF is parametrized using two Gaussians with a common mean value. For $D^{+}_{s}\to K^{0}_{S}\pi^{+}$, we fix the ratio of widths and the fractional yields in two Gaussians fits due to a lower signal sensitivity. The values of the ratio and the fraction are obtained from the fit to $D^{+}\to K^{0}_{S}\pi^{+}$ mode and are consistent with fit results from the simulated events of signal modes. $K^{+}/\pi^{+}$ misidentified backgrounds appear in the way described earlier and their normalizations are fixed during the fits to mass distributions. Table \ref{CS_Yields} summarizes the extracted signal yields from the fits to data and corresponding signal efficiencies from the simulated events of signal modes.

Various contributions to the systematic uncertainties for the branching ratio measurements are summarized in Table \ref{CS_Systematics}. The efficiency differences in particle identification are estimated and the central values of our ratio measurements are corrected. Uncertainties of such corrections are included in the systematics estimate and are found to be 0.90\% of the measured ratios. We refit data with varying bin size of histograms, different fit interval, and changes in the combinatorial background PDF. We estimate 0.24\% and 0.87\% of the measured ratio as systematic uncertainties due to variations in fit methods for $D^{+}$ and $D^{+}_{s}$ modes, respectively. Particle identification and its normalization effects of the $K^{+}/\pi^{+}$ misidentified background yields in fits are also estimated and found to be 0.16\% and 0.62\% of the measured ratio for $D^{+}$ and $D^{+}_{s}$ modes, respectively. Finally, systematic effects due to extra constraints in the $D^{+}_{s}\to K^{0}_{S}\pi^{+}$ signal PDF are estimated by re-fitting the data allowing the fixed parameters to change within their one standard deviation. This gives a systematic effect corresponding to 0.37\% of the measured ratio in $D^{+}_{s}$ decay modes. 

With the signal efficiencies and the corrections due to particle identification efficiency differences, we find branching ratios to be $\mathcal{B}(D^{+}\to K^{0}_{S}K^{+})/\mathcal{B}(D^{+}\to K^{0}_{S}\pi^{+}) = 0.1899\pm 0.0011\pm 0.0018$ and $\mathcal{B}(D^{+}_{s}\to K^{+}\pi^{+})/\mathcal{B}(D^{+}_{s}\to K^{+}_{S}K^{+}) = 0.0803\pm 0.0024\pm 0.0012$  where the first uncertainties are statistical and the second are systematic. These are the most precise measurements up to date and are compared to the present world average values in Table \ref{CS_Result}.

\section{Conclusions}

Using 605fb$^{-1}$ of data collected with the Belle detector at the KEKB asymmetric-energy $e^{+}e^{-}$ collider we have observed for the first time the decay $D^{+}_{s}\to K^{+}K^{+}\pi^{-}$ with a statistical significance of 9.1 standard deviations. This is the first DCS decay mode of the $D^{+}_{s}$ meson. The branching ratio with respect to the CF decay is $(0.229\pm 0.028\pm 0.012)\%$, where the first uncertainty is statistical and the second is systematic. We have also determined the $D^{+}$ DCS decay branching ratio, $\mathcal{B}(D^{+}\to K^{+}\pi^{+}\pi^{-})/\mathcal{B}(D^{+}\to K^{-}\pi^{+}\pi^{+}) = (0.569\pm 0.018\pm 0.014)\%$. with significantly improved precision compared to the current world average \cite{PLB667_1}. We find the product of the two relative branching ratios, $\frac{\mathcal{B}(D^{+}_{s}\to K^{+}K^{+}\pi^{-})}{\mathcal{B}(D^{+}_{s}\to K^{+}K^{-}\pi^{+})}\frac{\mathcal{B}(D^{+}\to K^{+}\pi^{+}\pi^{-})}{\mathcal{B}(D^{+}\to K^{-}\pi^{+}\pi^{+})}$, to be $(1.57\pm 0.21)\tan^{8}\theta_{C}$. This is consistent with SU(3) flavor symmetry within 3 standard deviations; note that the effect of (different) resonant intermediate states is not taken into account in the prediction \cite{NPBPS115_117}. An amplitude analysis on a larger data sample may allow a more precise test of SU(3) flavor symmetry to be performed.

\begin{table}[t]
\begin{center}
\caption{Branching ratios for $D^{+}$ and $D^{+}_{s}$ (preliminary), and comparisons with previous measurements. The uncertainties shown include the statistical and systematic uncertainties of our results.}
\begin{tabular}{lcc}
\hline\hline
Branching Ratio & Belle exp. & World average \cite{PLB667_1} \\
\hline
$\frac{\mathcal{B}(D^{+}\to K^{0}_{S}K^{+})}{\mathcal{B}(D^{+}\to K^{0}_{S}\pi^{+})}$ & $(19.0\pm 0.2)\%$ & $(20.6\pm 1.4)\%$ \\
$\frac{\mathcal{B}(D^{+}_{s}\to K^{0}_{S}\pi^{+})}{\mathcal{B}(D^{+}_{s}\to K^{0}_{S}K^{+})}$ & $(8.0\pm 0.3)\%$ & $(8.4\pm 0.9)\%$ \\
\hline\hline
\end{tabular}
\label{CS_Result}
\end{center}
\end{table}

We also have measured branching ratios with respect to the CF mode and values are $\mathcal{B}(D^{+}\to K^{0}_{S}K^{+})/\mathcal{B}(D^{+}\to K^{0}_{S}\pi^{+}) = 0.1899\pm 0.0011\pm 0.0018$ and $\mathcal{B}(D^{+}_{s}\to K^{0}_{S}\pi^{+})/\mathcal{B}(D^{+}_{s}\to K^{0}_{S}K^{+}) = 0.0803\pm 0.0024\pm 0.0012$. Using the world average values of CF decay modes \cite{PLB667_1}, we obtain branching fraction values of $\mathcal{B}(D^{+}\to K^{0}_{S}K^{+}) = (2.75\pm 0.08)\times 10^{-3}$ and $\mathcal{B}(D^{+}_{s}\to K^{0}_{S}\pi^{+}) = (1.20\pm 0.09)\times 10^{-3}$. These are consistent with the present world average and best measurements up to date. The ratio $\mathcal{B}(D^{+}\to K^{0}_{S}K^{+})/\mathcal{B}(D^{+}\to K^{0}_{S}\pi^{+}) = 2.29\pm 0.18$, which implies possible flavor SU(3) breaking or different final-state interaction between $D^{+}$ and $D^{+}_{s}$ decays.
\newline

%%%%%%%%%%%%%%%%%%%%%%%%%%%%%%%%%%

\begin{acknowledgments}
We thank the KEKB group for excellent operation of the
accelerator, the KEK cryogenics group for efficient solenoid
operations, and the KEK computer group and
the NII for valuable computing and SINET3 network support.  
We acknowledge support from MEXT, JSPS and Nagoya's TLPRC (Japan);
ARC and DIISR (Australia); NSFC (China); 
DST (India); MEST, KOSEF, KRF (Korea); MNiSW (Poland); 
MES and RFAAE (Russia); ARRS (Slovenia); SNSF (Switzerland); 
NSC and MOE (Taiwan); and DOE (USA).
\end{acknowledgments}

\bigskip % extra skip inserted

\end{document}